# History, mathematics and physics: two cases of the inverse square law

Isobel Falconer, School of Mathematics & Statistics, University of St Andrews

## 1 Introduction

How did mathematicians of the past relate to history? What can we learn from their experience? It is a truism to say that all people are embedded in history. It impacts on them in ways that they may or may not realise. At the two extremes, some may be completely oblivious; others may deliberately set out to change the historical narrative and the course of history.

In this chapter, I will explore these questions through two case studies. Both are of eminent British mathematicians/physicists in the late 1800s and their handling of an inverse square law of force. Both case studies show ways in which these men were blind to the history of their subject, conscious of it but limited by it, or deliberately trying to use it. I am grateful that in this book, Lizhen Ji has given me the opportunity to explore these questions, and reflect on their implications for today.

First, I look at William Thomson's and James Clerk Maxwell's attempts to establish the inverse square law of electrostatics in the 1850s to 1870s. I ask the question, why did Maxwell devote the last years of his life to editing the 100-year-old electrical researches of Henry Cavendish and repeating Cavendish's experimental verification of the inverse square law of electrostatic repulsion?

Second, I discuss John Henry Poynting and Newton's law of gravitation in the 1880s and 1890s, asking why Poynting preferred to describe his experiments as measurements of the mean density of the Earth rather than of the gravitational constant (Falconer 1999a).

By looking at case studies around a simple mathematical model of the natural world, we can investigate three distinct historical strands: the role of changing mathematical forms, the role of the historical accumulation of experimental evidence, and the changing dynamics of the discipline. The case studies are of the same era as that singled out by Henri Poincaré when he lamented a change in mathematics as a discipline, 'Formerly, when a new function was invented, it was in view of some practical end. To-day they are invented on purpose to show our ancestors' reasonings at fault, and we shall never get anything more than that out of them' (Poincaré 1914 p125). But the case studies belong to a different part of the discipline, one that was increasingly divorced from pure mathematics, and so seem likely to give different answers to my initial questions from studies of pure mathematics.

Thus, it is relevant to establish at the outset how the main actors in these stories regarded themselves. All were highly trained mathematicians, graduates of Cambridge University's gruelling Mathematical Tripos which dominated British mathematics in the 1800s. Thomson and Maxwell were second wranglers (2$^{nd}$ top) in 1845 and 1854 respectively, Poynting was third wrangler (1876). However, all three went on to hold chairs in Natural Philosophy, or its more recent incarnation, Experimental Physics. For them mathematics and physical understanding were unified, as two sides of the same coin. They developed new mathematics as they needed it to increase their physical understanding. Thomson devised the method of electrical images through which electrostatic problems were transformed into a more tractable form by inversion in a sphere (Lutzen 2011). He subsequently developed the mathematics of vortices to understand magneto-optical rotation and to model atoms (Harman 1982, Falconer 2019). Maxwell, famously, built on Faraday's and Thomson's work, and on the quaternions of William Rowan Hamilton to model his new concept of an electromagnetic field, contributing to the development of vector calculus methods and nomenclature along the way (Crowe 1984). He was a major, though informal, contributor to the knot theory being developed by his friend, Peter Guthrie Tait (Epple 1998). Poynting is best known for his mathematical analysis of the flow of energy in an electromagnetic field and for deriving the 'Poynting vector' which describes it (Pellosi & Seleri 2017).[1]

## 2 Thomson, Maxwell and the inverse square law of electrostatics

In 1877 James Clerk Maxwell and his student Donald McAlister refined Henry Cavendish's 1773 null experiment demonstrating the absence of electric charge (electrification) inside a charged conductor.[2] Such absence of charge was a mathematical prediction of the inverse square law of electrostatics, and both Cavendish and Maxwell took the experiment as verifying the law.

However, in his famous *Treatise on Electricity and Magnetism* Maxwell had already expressed absolute conviction in the law, based on results of Michael Faraday (Maxwell 1873 p75). So why did he spend his time editing Cavendish's unpublished papers and replicating the experiment?

The usual, short, answer is that Maxwell owed a duty to the 7th Duke of Devonshire, head of the Cavendish family, who had recently paid for building a physics laboratory at the University of Cambridge. Maxwell had been appointed to lead the laboratory as the first 'Cavendish Professor of Experimental Physics at Cambridge (Harman 2002 p11, Whitaker 2013 p116).

A more thorough investigation shows that this is a story of attempts to establish the superiority of mathematical over experimental approaches to physics (Falconer 2017a). It shows that even respected mathematical scientists can be so oblivious to the historical tradition in which they are working that they do not notice weaknesses in their arguments. When Maxwell and Thomson do consciously use historical cases, they are guilty of misrepresentation for the sake of telling a persuasive story.

The inverse square law of electrostatics states that the repulsive force $F_E$ between two like electrified bodies is given by

$$F_E \propto \frac{e_1 e_2}{r^2}$$

where $e_1$ , $e_2$ are the values of the electrification of the two bodies and $r$ is the distance between them. It was first suggested by Joseph Priestley in 1767 by analogy to Newton's law of gravitation, and for many it carried conviction because of the success of Newton's law. Both laws attempted only to describe the observed phenomena mathematically without enquiring into the underlying force mechanism. Thus, for around 150 years, since Newton, an assumption of action at a distance prevailed, and was sufficient for the static case, but by the mid-1800s it was proving insufficient for dynamic situations, as recognised by Maxwell's field theory and Weber's rival electrodynamic theory (Weber, 1893). The inverse square law was fundamental to both, but while Weber generalised it by introducing velocity into an action-at-a-distance force, Maxwell introduced fields, adopting Gauss's generalisation of the law.

[1] The Russian physicist and mathematican Nikolay Umov, and the British physicist Oliver Heaviside, independently developed similar ideas.

[2] Throughout this section on Thomson and Maxwell, I refer to 'electrification' or 'electrified bodies' where today we would talk of 'charge'. Thomson and Maxwell did not use a concept of discrete electric charges at this time. For them 'electrification' represented some state of the continuous electromagnetic ether or field, though they occasionally talked of 'discharge' of electrified bodies.

According to traditional accounts, the inverse square law of electrostatics was proven experimentally by Coulomb's 1785 torsion balance experiments conducted nearly 100 years earlier than Maxwell's (Coulomb 1884); it is often termed 'Coulomb's Law'. However, Coulomb's experiment comprised only three data points which were all close together. This left plenty of scope for measurements in different conditions to produce different laws (Heering 1992).

But, contrary to the traditional account, perceptions of the status of the inverse square law varied widely and were changing rapidly prior to 1870. This is evidenced by an examination of British publications on electrostatics in the mid-1800s (Falconer 2017a). From the mid-1830s, two opposed approaches developed. The first, led by one of Britain's most prestigious electrical scientists, Sir William Snow Harris, was an inductive, experimental one. Harris, and his follower, the telegraph engineer Frederick Charles Webb, held that the inverse square was not fundamental and was obeyed only in some circumstances. They based their claim on Coulomb's limited experiments and their own direct quantitative measurements. They held that Coulomb's results represented the composition of a direct inverse relation with electrical induction effects and held only for suitable arrangements of conductors. Harris promoted this view until his death in 1866 and in his posthumous *Frictional Electricity* (Harris 1867).

The second approach, a mathematical deductive one, was spearheaded by William Thomson. In 1845 Joseph Liouville asked Thomson to clarify the issues raised by Harris' challenges to the laws of electrostatics. At the time, Thomson was recently graduated, yet to make his name, and very much at a social disadvantage to Harris who had been knighted for his services to electrical science (James 2004). But Harris's experimental approach was anathema to Thomson. He had recently read Green's little-known paper on the uses of potential theory (Green 1828). This equipped him with methods for mathematising electrostatic theory using differential equations for measurable macroscopic quantities, thus avoiding many of the contradictions and problems introduced by microscopic hypotheses about electricity. But the validity of applying Green's theorem, and Thomson's own new method of electrical images, to electrostatics depended upon the truth of the inverse square law.

Thomson was impelled to defend the law vigorously, attributing Harris' results to disturbing influences and unjustified generalization (Thomson 1872 p21). Thomson viewed the inverse square law as a mathematical truth; the beauty and power of the mathematics drove his interpretation of experimental results. He saw at stake both the mathematical power of potential theory and the status of mathematical scientists such as himself *vis a vis* experimental inductive scientists such as Harris. Over the next 30 years, as he developed his arguments, a nationalistic theme also emerged as Thomson, and Maxwell following him, sought to persuade a British audience by creating a story of British proof of the inverse square law.

Thomson's first response to Liouville's request already showed the importance he attached to a deductive approach which prioritised mathematics. In 1845 he made clear that the form of the law was dictated by mathematics, and hence must be true:

> '...[Coulomb] has thus arrived by direct measurement at the law, *which we know by a mathematical demonstration*, founded upon independent experiments, *to be the rigorous law of nature for electrical action*,' (Thomson 1872 pp24-25, my emphasis).

He referenced two Cambridge-educated mathematicians, Robert Murphy (1833) and John Pratt (1836), for the mathematical demonstration, either ignorant of, or ignoring, Laplace's much earlier demonstration (Laplace 1799 p143). At this point he was vague about the nature of both the mathematical demonstration and the 'independent experiments', but in 1854 he added a note specifying these.

> 'Cavendish demonstrates mathematically[3] that if the law of force be any other than the inverse square of the distance, electricity could not rest in equilibrium on the surface of a conductor. But experiment has shown that electricity does rest at the surface of a conductor. Hence the law of force must be the inverse square of the distance' (Thomson 1872 p24 fn).

We can see framework of Thomson's basic argument, which did not change through the years:

1. An inverse square law predicts mathematically that no electrification would exist inside a closed conductor, it would all reside on the surface (Cavendish and Thomson's reasoning is described below):
2. Numerous scientists have observed the absence of electrification inside a closed conductor qualitatively:
3. This proves the law.

Note that in the quote above, Thomson named the British aristocrat Henry Cavendish, who was far better known for his work on gravitation, but the, mainly French, electricians such as Coulomb, Jean-Baptiste Biot, and Jean-Gustave Bourbouze, whose experiments he referred to, remained anonymous. He was unaware when he wrote this in 1854 that Cavendish had performed a careful, quantitative, experiment, but he did now know that Harris possessed a large collection of Cavendish manuscripts on electrical science; Thomson was convinced that they were extremely important. He began a campaign to get them published (Falconer 2017b).

By around 1870, the situation had changed, and Thomson's mathematical deductive approach was gaining ground following Harris' death in 1867. Thomson was by now Professor of Natural Philosophy at Glasgow and had been knighted himself in 1866 for his work on the transatlantic telegraph cable. But he evidently did not feel that the mathematical approach was sufficiently established. He enlisted Maxwell's help in locating and publishing Cavendish's papers. Maxwell's assistance became more effective in 1871 when he was appointed as the first Professor of Experimental Physics at Cambridge, heading a laboratory funded by Cavendish's relative the 7th Duke of Devonshire. Maxwell eventually received the manuscripts in1874 and started editing them for publication, repeating experiments he deemed important (Falconer 2017b). Among these was Cavendish's inverse square law experiment, which no-one except Harris had known of previously.

---

[3] Cavendish's 'mathematical demonstration' was a reasoned argument but did not use symbols, equations or explicit geometry.

Although Thomson and Maxwell disagreed about the processes underlying electromagnetism, the inverse square law was foundational to both. They shared beliefs in a physics based on the concept of energy, and the prime importance of precise measurement and mathematical law. They now sought to establish the law of electrostatics securely and beyond doubt, along the lines of Thomson's proof sketched above. Since this method was a an indirect one, dependent on mathematical deduction, it carried with it the implication that electrical scientists needed to be mathematically educated in order to undertake, or even to understand, proper scientific method. As we shall see, Thomson and Maxwell's campaign proceeded through precise measurement, a gap between mathematical and experimental models, and an inaccurate reconstruction of history.

Cavendish, in his 1771 mathematical demonstration referred to by Thomson (1872 p24 fn) had assumed an exact inverse square law and used a one-fluid model of electricity to reason that the electricity on a closed spherical conductor resided at the surface, with no electricity inside (1771, pp592-4). Consider an electrified spherical conducting shell. By symmetry, and since the electricity can move around on the conductor, it will distribute itself uniformly over the surface.[4] Then (in modern terms that Cavendish did not use) consider the electric force at any interior point, P, by drawing a cone of infinitesimal solid angle through P that cuts the sphere on both near and far sides (Figure 1). For an inverse square law, the force at P is $\frac{\rho dS_n}{r_n^2} - \frac{\rho dS_f}{r_f^2}$ where $\rho$ is the surface density of electricity, $dS_n$, $dS_f$ are the infinitesimal surface areas of the sphere cut by the cone on near and far sides, and $r_n$, $r_f$ are the distances from the point to the surface of the sphere.

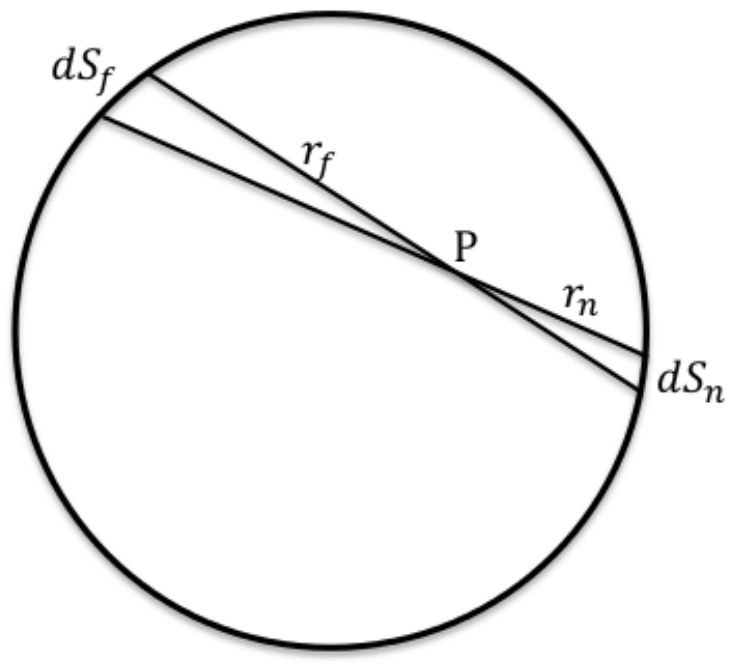

*Figure 1. The geometry of an inverse square force on a point inside a spherical conducting shell.*

By geometry, the infinitesimal areas $dS_n$, $dS_f$ increase as the square of $r_n$, $r_f$, so write them as $dS_n = \alpha r_n^2$ and $dS_f = \alpha r_f^2$. Hence the force at P becomes $\frac{\rho \alpha r_n^2}{r_n^2} - \frac{\rho \alpha r_f^2}{r_f^2} = 0$, since the $r_n^2$, $r_f^2$ and terms cancel out. This result holds whatever the direction of the cone, hence the total resultant force is zero everywhere inside the shell. If the microscopic force obeys a power law that differs from inverse square by a small amount, the resultant will be directed towards or away from the centre. Then, argued Cavendish, suppose a globe is placed inside the outer shell, connected to it by a wire, and consider the free electric fluid in the wire. If the inverse power differs from two, the fluid will move towards (or away from) the centre, leaving the inner globe with excess (or deprived of) fluid. Either way, unless the power is exactly two, the inner globe will be electrified.

---

[4] Cavendish referred to the similar situation of a particle inside a massive shell, considered by Newton in proposition 70 of Book 1 of the *Principia*

Maxwell now found from the manuscripts that ten years ahead of Coulomb, in 1772-3, Cavendish had tested his prediction experimentally, concluding that the negative exponent could not differ from two by more than about 1/50 (Maxwell 1879 pp111-112). Establishing the superiority of Cavendish's mathematical-deductive method, and his priority over Coulomb, seems to have been Maxwell and McAlister's main motivation in improving Cavendish's experiment to a sensitivity of 1/21600. They published their experiment shortly before Maxwell's death, in *The Electrical Researches of Henry Cavendish* (Maxwell 1879 pp417-422), and in the posthumous second edition of Maxwell's *Treatise* (Maxwell 1881 pp77-82).

A logical problem with Thomson's proof, sketched above, is that although an inverse square law predicts, with absolute certainty, that all the electricity resides on the outside of a closed conductor, the converse is not necessarily true. Unlike Thomson, Maxwell was well aware of this problem of reverse inference, and raised it in his draft account of Cavendish's methodology.

> It is obvious that the mere agreement between experiment and the deductions from an hypothesis cannot prove that hypothesis to be true unless it can also be shown that no other hypothesis will agree with the experiments (Harman 2002 pp541-542).

Maxwell had already given the required mathematical demonstration, that no other hypothesis would agree with the null result, in his 1873 *Treatise on Electricity and Magnetism*, before knowing of Cavendish's experiment, His demonstration derived from the Cambridge tradition within which he was working and this had repercussions for his repetition of Cavendish.

Although, like Thomson in 1845 and 1854, Maxwell cited Pratt (1836) for the demonstration, his version was actually due to his friend, Peter Guthrie Tait, who outlined it on a postcard written in June 1871(Harman 1995 p650). Maxwell (following Tait) began by defining a spherically symmetric potential function due to a unit of electricity at a distance $r$ as $f(r)\ /\ r$. Then he posited a spherical shell of radius *a* and derived the potential $V$ at a point P a distance $p$ from the centre by integrating over the surface of the shell. He obtained $V = 2\pi\sigma\{f(a+p) + f(a-p)\}$, where $\sigma$ is the surface density of electricity. Using the experimental observation that there is no electricity inside the shell, and hence that $V$ is constant for all values of *p* < *a*, he differentiated $V$ with respect to $p$ to obtain $f'(r)$ in the first of the three lines from which he derived the inverse square law:

$$0 = f'(a+p) - f'(a-p) \qquad (1)$$

Since *a* and *p* are independent, (2)

$$f'(r) = C, \text{ a constant} \quad (1873a, p. 76). \qquad (3)$$

Integrating $f'(r)$ to obtain $f(r) = Cr + C'$, substituting into the expression for the potential function, and differentiating the potential function to give the electric force, gave the force as proportional to $\frac{1}{r^2}$.

To modern eyes, a problem with (3) is that it is true only if the first line applies *for all a*. This can be seen by starting with (1), which requires that $f'(a-p) = f'(a+p)$. Then, if $a$ is constant, any function $f'(r)$ which is symmetric about $a$ satisfies the requirement as *p* varies independently. So $f'(r) = C$ works, but so does $f'(r) = cos(r-a)$ or $f'(r) = (r-a)^2$ (or a multitude of other even functions). Only If $a$ varies as well as $p$, does the conclusion that $f'(r) = C$ hold.

Maxwell had inherited the tradition set up by his Cambridge antecedents, Murphy, Pratt, Thomson, and Tait. Murphy and Pratt had used series expansions in their demonstrations, which Thomson accepted and promoted. None of them mentioned the condition that *a* must vary and, as we have seen, neither did Tait. They may have assumed implicitly that *a*, a variable, would vary over all possible values. All the while this remained a textbook demonstration, that was not an issue. But it became an issue for Maxwell when he repeated Cavendish's experiment using a conducting shell with a *single* value of *a* and wished to claim it as verification. This would introduce a gap between the implied logic of his mathematical demonstration and the logic of his experiment.

Maxwell had two options for avoiding this gap in his edition of Cavendish's electrical researches, and the second edition of his *Treatise on Electricity and Magnetism*. He could assume, as Cavendish had done, that the force law was an inverse power (Maxwell 1879 p422, Maxwell 1881 pp 81-82). With this condition, Tait's demonstration held for any single value of *a*. But Maxwell wished to relax the restriction so that he could claim (as quoted above) that *no* other hypothesis agreed with experiment. Alternatively, he could stick to a more general demonstration – this time he chose that due to Pierre Simon Laplace (1799) rather than Tait's – but this would entail conducting multiple experiments with shells of many different radii. Unlike Tait, Laplace was working in a more precise French tradition and was explicit that *a* had to vary over all possible values if the result was to hold.[5]

Maxwell's desire to use a general mathematical demonstration while at the same time hailing Cavendish's as *the* authoritative experimental verification of the law, led to a very confused argument (Maxwell 1881 pp81-82). He included the restriction to an inverse power law in his account of the theory of Cavendish's experiment. Next, he asserted that, 'Laplace gave the first demonstration that no function of the distance except the inverse square satisfies the condition that a uniform spherical shell exerts no force on a particle within it.' He gave a summary of Laplace's general demonstration but glossed Laplace's 'since this relation holds for all values of' with the less precise, 'this equation is generally true' (Laplace 1799, p143, translated in Todhunter 1873 pp181-182; Maxwell 1881 p81). Finally, he invoked evidence from other experiments when he stated that although Cavendish's assumption 'may appear less general than that of Laplace', it was the only one consistent with the observed electrification of similarly shaped conductors, thus undermining his previously implied claim that Cavendish's experiment was sufficient.

The 20th-century philosopher, Jonathan Dorling, has described Maxwell as 'disingenuous' for using only one sphere and not mentioning the implications of Laplace's mathematical demonstration (Dorling 1974 p336). Rather than disingenuity, I suggest that the episode shows the extent to which Maxwell was embedded in history. Dorling viewed the experiment as an isolated one, separate from two traditions within which Maxwell was working. First, and most obviously, Maxwell was aware of the accumulating qualitative observations of the absence of electrification inside closed conductors of many shapes and sizes, from which some experimental induction might be allowed even if those experiments were rough and ready.

---

[5] In 1873 the Cambridge mathematician Isaac Todhunter published an account of Laplace's work. This is probably the source of Maxwell's knowledge of Laplace, and accounts for the change between the first and second editions of his *Treatise*. Like Laplace, Todhunter *did* include the condition 'for all values of *a*' in his demonstration (Todhunter 1873 pp181-182).

Second, as discussed above, he inherited a historical Cambridge mathematical tradition that did not state its assumptions explicitly, rendering it easy to overlook the problems these could cause for experimental design.

But as well as being uncritically embedded in history, Maxwell and Thomson constructed a narrative of proof of the inverse square law that would appeal to British scientists and attract support for their mathematical-deductive approach to physics. Although they did not explicitly badge as ‘history’ their systematic account of the steps they deemed important in establishing the law, it amounted to an implicit history. Anyone reading it would receive the impression that only the British had made significant contributions to the development of electrical science. Thomson took the first steps in creating a narrative that focused on Henry Cavendish and Michael Faraday. He had had his first brief sight of the Cavendish manuscripts in 1849 when they were in Harris’s hands (Falconer 2017b). Two years later Cavendish’s biographer, George Wilson, recorded Thomson’s view that the papers contained, ‘descriptions of excessively ingenious experiments leading to important quantitative results,’ and that they should be published (Wilson 1851 p469). We have already seen how Thomson pointed to Henry Cavendish, but not the French electrical scientists, by name in his argument for the truth of the inverse square law in 1854. In the same year he told James David Forbes, Professor of Natural Philosophy at Edinburgh, that he was, ‘disposed to regard Cavendish as the founder of the Mathematical Theory of Electricity,’ again urging publication of the manuscripts (Thomson 1854). On his advice, Forbes prioritised Cavendish’s contribution over Coulomb’s in his widely read history of the progress of mathematical and physical science from 1775-1850, written for the eighth edition of the *Encyclopaedia Britannica* (Falconer 2021 p56).

In similar vein, Maxwell, in his *Treatise* in 1873, used the observation of absence of electrification inside a conductor as proof of the physical validity of the theory. But unlike Thomson he did not refer vaguely to ‘experiment’. Instead, he explicitly named Faraday as the significant experimenter. Faraday was, by this time, Britain’s most prominent electrical scientist, with a high reputation for trustworthiness and careful experiments, and Maxwell leveraged this reputation to support his own claims. He acknowledged Coulomb’s 1785 torsion balance experiments as establishing the inverse square law of electrostatics, merely to dismiss them as demonstrating it only to a rough approximation. Instead, he cited Faraday’s observation that an electrified body, touched to the inside of a conducting vessel, transfers *all* its electricity to the outside surface, as ‘far more conclusive than any measurements of electrical forces can be’ (Maxwell 1873 p75).

Once again, there is a problem. This time it is that Maxwell misrepresented Faraday’s experiments. An observational inverse square law is a *necessary* condition for potential theory to be of any use in electrostatics. To bolster the observational evidence, Maxwell shifted the evidential context of Faraday’s experiments and distorted their description. Earlier in the *Treatise*, he had already described the experiments in their proper context of electrification by induction – the context in which Faraday had made them and intended them as evidence (Maxwell 1873 pp31-32; Faraday 1843). Maxwell now invited his readers to take the same experiments as conclusive evidence for the inverse square law (Maxwell 1873 p75). However, If we go back and look at Faraday’s original paper, we find that he conducted the experiment as an illustration of his own ideas about electrostatic induction (Faraday 1843). Faraday made no mention of the law of force, and he described the outer conductor as an open ‘pail’; his diagram showed it as a cylinder and wide open at the top. To claim these experiments as evidence for the inverse square law, Maxwell had to misrepresent them textually and graphically. He repeatedly described Faraday's containers as closed, 'completely surround[ing]' the electrified body (Maxwell 1873 pp31, 75). His diagram showed it with a lid that could be fully closed and with rounded ends so that it looked more nearly spherical (Figure 2) (Maxwell 1873 p31). That he closed the container in both text and diagram suggests that he was aware that, mathematically, Faraday's experiments provided no evidence for the inverse square law, since the theoretical prediction applied only to *closed* containers.[6] He was writing Faraday into his account of proof of the law, even though the experiments he cited did not warrant it.

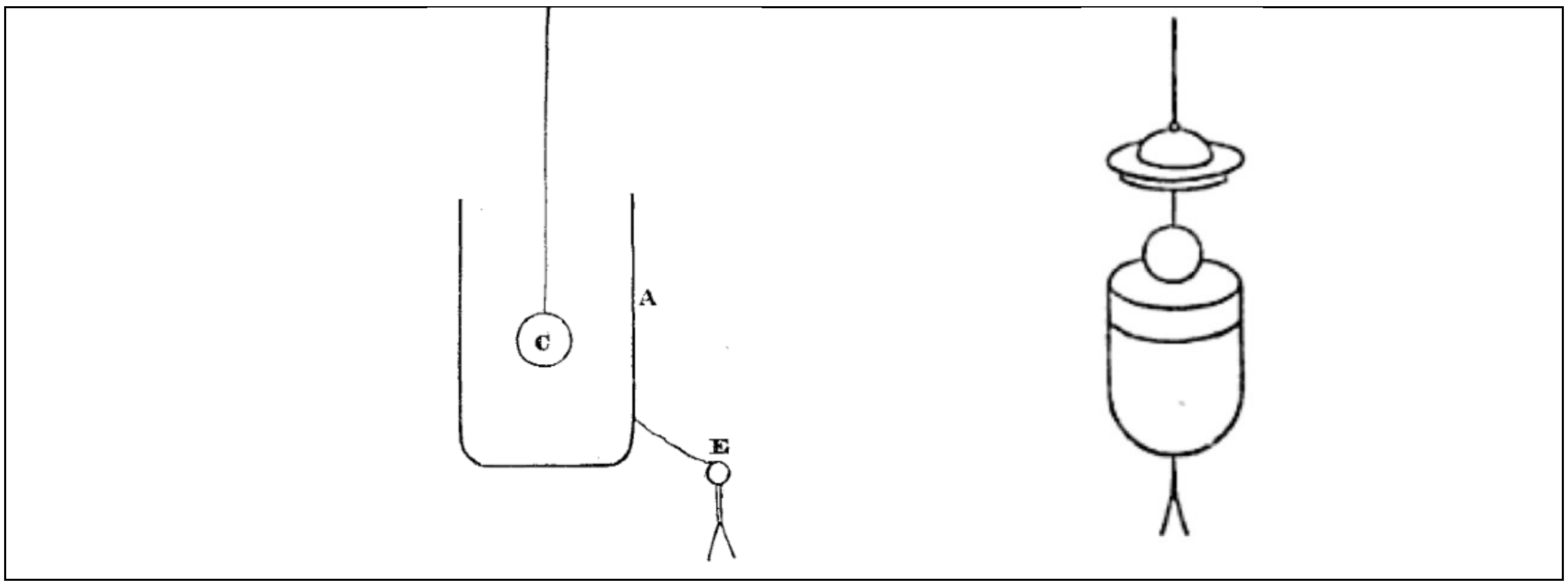

*Figure 2. Faraday's diagram of his apparatus for electrification by induction (left) (1843 p200), and Maxwell's diagram of Faraday's apparatus (right) (1873 p31).*

Three years later, Thomson was even more emphatic than Maxwell that Cavendish and Faraday deserved the credit for the inverse square law. Speaking in 1876, before he knew of Cavendish's experiment, he held that,

> 'The other main method of experimenting... is illustrated also by Cavendish's [theoretical] writings, that is the seeking for a zero.... It was left for Faraday to make with accuracy the concluding experiment which crowned Cavendish's theory. Faraday [1843] found by the most thoroughly searching investigation that the electrical force ... was zero.... Therefore the law of force varies with the inverse square of the distance. This result was obtained with far less searching accuracy by Coulomb and Robinson[sic], because their method did not admit of the same searching accuracy.' (Thomson 1876 p241)

Despite Thomson's claims of accuracy, such precision does not seem to have been a concern of Faraday's. His paper described only a visual indication using a 'delicate' gold leaf electroscope to test whether the interior ball was electrified. 'If C [the ball] be made to touch the bottom of A [the cylinder] ... C, upon being withdrawn and examined, is found perfectly discharged.' Notably, he gave no indication of how he had examined the electrification of the ball, finding it to be zero. He clearly thought this unnecessary, for he continued, 'These are all well-known and recognised actions', presumably alluding to the many (French) experiments that Thomson and Maxwell had dismissed (Faraday, 1843 p201).

---

[6] Other experiments of Faraday's, such as the 'Faraday cage' experiment, would have served Maxwell's argument better, but they are not ones he used here.

The final stage in Thomson and Maxwell's implied history came in the second, posthumous, edition of Maxwell's *Treatise on Electricity and Magnetism*. By now he had repeated Cavendish's experiment so, after dismissing the value of Coulomb's experiments, and a brief mention of Faraday's as superior, he gave all the credit to Cavendish, explaining both the experiment and the theory, and his own repetition of it, in some detail (Maxwell 1881, pp76-82).

Unlike Faraday, Maxwell and McAlister did assess the precision of their zero measurement. For them, assessing this precision was the whole point of the experiment as it set the limits on how much the exponent of distance could differ from two. However, Ronald Laymon (1994) has pointed out that they, and Cavendish before them, assessed the precision of their measurements using instruments that relied on the inverse square law for their operation! This provides further insight into the difficulty of recognising the assumptions of one's own tradition, and the impact this may have on designing experiments.

Through mismatch between mathematical demonstration and experimental design, and a distorted implicit history, Thomson and Maxwell defended the inverse square law vigorously. But their defences always depended ultimately on the conviction that the law was true. All were predicated on the general knowledge that there is, indeed, no electrification inside a closed conductor. The power of the mathematics available to them, and their investment in an electrical programme based on it, gave them the confidence and the incentive to overlook the gaps and mensurational weakness of Faraday's experiments as evidence for the inverse square law. They leveraged his high reputation for their own ends in asserting the validity of the law and, by implication, of their own mathematical-deductive approach to physics.

## 3 Poynting and the gravitational constant

The subject for Cambridge University's Adams Prize for 1893 was 'The Methods of determining the absolute and relative value of gravitation and the mean density of the earth.' John Henry Poynting received the prize for his essay which included both a history of methods and measurements of the mean density of the earth (MDE), and an account of his own 13 years-worth of measurements made using a common balance (Poynting 1894a). Throughout, and for at least the next six years, Poynting consistently preferred to interpret his experiments as measurements of MDE, rather than of the gravitational constant. Why?

Today, we attach significance to $G$ as a fundamental and universal constant - so Poynting's preference looks perverse. But this overlooks the fact that the idea of a universal constant had entered gravitation models only 20 years earlier and was still very new. A quick view of the sketch that follows, of the history of gravitation experiments, might suggest that Poynting was simply a reactionary conservative. But the more probing investigation that follows the sketch shows that he had good reason. Along the way it illuminates one way in which mathematical form may affect physicists' concepts and research questions about the natural world.

The Adams prize topic reflected what Poynting termed a recent 'epidemic of earth-weighing' (Poynting 1894b p627). After surveying nineteen experiments, dating from 1735 to 1894, Poynting was confident that the results of this epidemic were converging towards a steady mean of 5.5 g $cm^{-3}$ for the density of the earth (Figure 3). This corresponded to approximately $7 \times 10^{-8}$ $cm^3$ $g^{-1}$ $s^{-2}$ for $G$, the 'constant of attraction' – a relatively new concept at the time. Today, but historically not until 1884, $G$ is known as the 'gravitational constant' that appears in the way we are accustomed to expressing the inverse square law of gravitation:

$$F_{12} = G\frac{m_1 m_2}{{r_{12}}^2},$$

where $F_{12}$ is the gravitational force between bodies 1 and 2, $m_1$ , $m_2$ are the masses of the two bodies, and $r_{12}$ is the distance between them. If one of those masses is the earth, and body 2 is on the surface of the earth, the expression becomes:

$$F_{E2} = G\frac{M_E m_2}{{r_E}^2},$$

where $F_{E2}$ is is the gravitational force of the earth on body 2, $M_E$ is the mass of the earth, and $R_E$ is the radius of the earth. From this, MDE can easily be calculated if the other parameters are known.

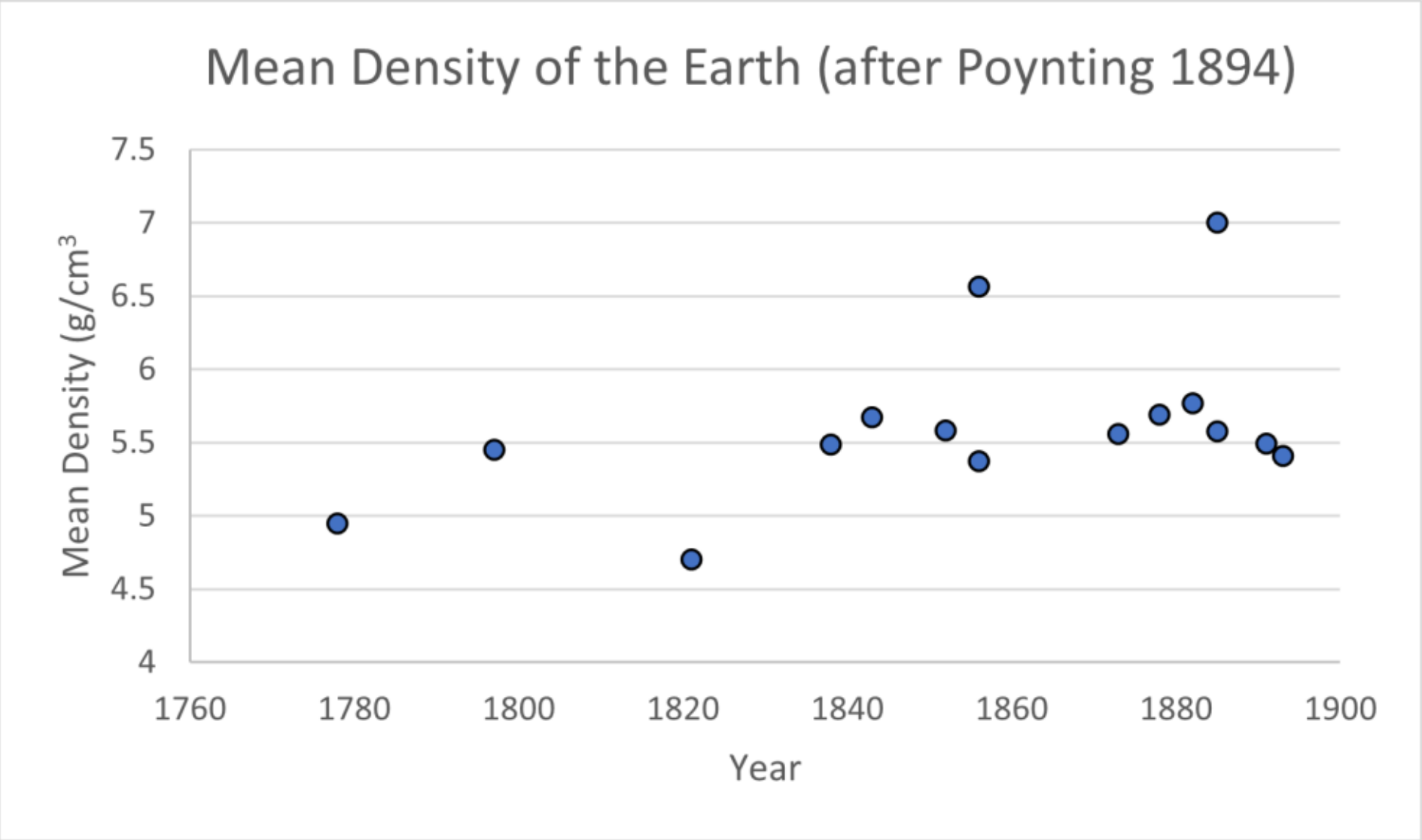

*Figure 3: Measurements of the mean density of the earth from 1778 to 1893 from the data in Poynting (1894)*[7]

Poynting was explicit about his preferred interpretation in his address to the Royal Institution in 1900. Talking of Charles Vernon Boys' recent measurements of $G$ he remarked:

> Professor Boys has almost indignantly disclaimed that he was engaged in any such purely local experiment as the determination of the mean density of the earth. He was working for the Universe, seeking the value of G, information which would be as useful on Mars or Jupiter or out in the stellar system as here on the earth. But perhaps we may this evening consent to be more parochial in our ideas, and express the results in terms of the mean density of the earth (Poynting 1902 p282)

This evidence of heated debate over the interpretation of the experiments is a clue that the history of the law of gravitation is not as straightforward as it is usually made out to be.

Poynting's (1894a) list of papers published since Newton announced his theory of gravitation 200 years earlier, shows that $G$ was a very recent innovation; until 1884 finding the mean density of the earth was the stated purpose of gravitation experiments.[8] But although at a cursory glance this might seem a regrettable hang over from a geocentric view that the Earth was by far the most important object in the universe, it was not. It was a necessary step in measuring the solar system and verifying the inverse square law.

---

[7] Bouguer's experiment of 1735 gave an inconclusive value and is not included in the data.
[8] The possibility that Poynting's was a biased list can be discounted. Other lists by the Canadian, Arthur Stanley Mackenzie (1899), and the German, Jonathan Zenneck (1903), give the same picture.

Newton himself had not used symbolic equations at all. His crucial statements of the three important claims of his law are couched in terms of proportions:

1. 'there is a power of gravity pertaining to all bodies, proportional to the several quantities of matter which they contain' i.e. gravity is *mutual* (if one uses symbols, it depends on the product $m_1 m_2$) and *universal* (applies to all bodies and at all distances) (Newton, *Principia* book 3, proposition 7, theorem 7; trans. Newton et al 1934 p414).
2. 'The force of gravity which tends to any one planet is inversely as the square of the distance of places from that planet's centre' (Newton *Principia*, book 3, proposition 5, theorem 5, corollary 2; trans. Newton et al 1934 p414).

Following Newton, 18th-century workers compared the attraction of one planet to that of another, expressing the results as ratios – no constant was needed, nor did one arise in the equations. Thus, the masses of the planets were known relative to one another; indeed, it was (and still is) impossible to find just from observation of planetary orbits, what their absolute masses are. In today's terms, but not those of the 1700s, only the product $GM$ (where $M$ is the planet's mass) could be obtained. Hence the question of a constant $G$ could not arise. In Boys' words:

> '… though we may know that twenty-eight suns would be required to make one Sirius ; that the sun is equal to 1048 Jupiters, that Jupiter is more than double all the rest of the solar system put together, or that the moon is 1/80 of the earth ; no observations of these bodies can ever tell us how many tons of matter go to make up any one of them.' (Boys 1894 p353)

To move forward, the mass of one planet needed to be measured in order to fix the masses of the others. The planet available for such independent measurement was the earth.

Several ways of achieving a measurement of the mass of the earth were possible; all relied on comparing the force of gravity at the surface of the earth on a body with that of a second body of known mass. This second body might be a mountain or strata of rock, whose mass could be calculated from the density of rock samples and surveyed measurements of the volume of the mountain or depth of strata. The earliest measurements of the mass or density of the earth were of this type. The first was by Bouguer in Peru in 1735 which gave inconclusive results. The next, by Nevil Maskelyne and Charles Hutton, used Schiehallien in the Scottish Highlands in 1784 and gave a value for the MDE of 4.5 times that of water (taken as 1 g cm$^{-3}$) (Poynting 1894).[9]

The alternative was laboratory measurement, using a manufactured body of known mass. In 1798 Henry Cavendish realised the first such laboratory comparison, using a torsion balance designed by John Mitchell. Despite frequent claims that this was the first laboratory measurement of *G*, Cavendish did no such thing (Clotfelter 1987, Falconer 1999b, Ducheyne 2011). He proceeded directly from the ratio of measured forces to obtain a value for MDE of 5.448 times that of water (Cavendish, 1796 p521), making no attempt to test the proportionality of the force to the two laboratory masses (Figure 4a). Measurements of MDE, both in the field and in the laboratory, using various means of measuring the attraction – pendulums, torsion

[9] Recalculated by Hutton in 1821 to give 4.95 times the density of water (Hutton 1821)

balances, common balances – continued for the next 100 years, increasing in number towards the end of the 19th century (see Figure 3).

But even now, with the mass of the earth known and the ability, in principle, to disaggregate the proportionality constant and the mass of a planet, the question of $G$ might not have arisen had the form of mathematics itself not begun to change consequent on Euler's introduction of functional notation in the mid-18th century (Euler 1748). In the notation of explicit functions, the dependent variable – in this case the gravitational force on a body – was typically written on the left-hand side of the equation with the requisite combination of independent variables on the right-hand side. This form lent itself to absolute, rather than relative, measurements; crucially, a coefficient of proportionality was required to satisfy dimensional requirements. Figure 4 compares the ratio and explicit formulations of Newton's law.

| a) 18th-century ratio form | b) Late 19th-century explicit form |
|---|---|
| $\frac{M_E}{{R_E}^2} : \frac{m_2}{{r_{12}}^2} = F_{E1} : F_{12}$<br>$V_E = \frac{4}{3}\pi {R_E}^3$<br>$\Delta_E = \frac{F_{E1}{R_E}^2 m_2}{F_{12}{r_{12}}^2 V_E}$<br>where $V_E$ is the volume of the earth, $\Delta_E$ is the mean density of the earth, and other parameters have already been defined | $F_{12} = G\frac{m_2 m_1}{{r_{12}}^2}$<br>$G = F_{12}\frac{{r_{12}}^2}{m_2 m_1}$<br>$F_{E2} = G\frac{M_E m_2}{{R_E}^2} = G\frac{\Delta_E m_2 V_E}{{R_E}^2} = g m_2$<br>$\Delta_E = g\frac{{R_E}^2}{G V_E}$<br>where $g$ is the local acceleration due to the earth's gravity |

*Figure 4. A comparison of the ratio and explicit forms of Newton's law, following Poynting's (1894a) account: a) the ratio form used in the 18th century, b) the explicit form used increasingly through the 19th century.*

It took time for this change to work its way through into astronomy, natural philosophy and physics. But we see it by 1798 for gravitation in Laplace's *Mécanique Céleste*:

$$\varphi = \frac{4\pi^2}{k^2}\frac{1}{r^2} ,$$

where $\varphi$ is the accelerating force between two bodies and $k$ is the proportionality constant (Laplace 1799 pp115-116). Over the next 100 years, the constant gradually acquired a greater meaning than purely to ensure homogeneous dimensionality and make the units come out right.

In Britain it took a while for the change in form to catch on, but Henry Pratt did so in his 1836 textbook, that Thomson and Maxwell had also referred to, and that was designed to introduce Cambridge students to Continental mathematics:

> 'Let $M$and $m$ be the masses of the two particles, $r$ their distance at the time $t$: then, if the unit of attraction be the attraction of a unit of mass at a unit of distance, the accelerating force produced in $M$ by the attraction of $m$=$m/r^2$ ' (Pratt 1836 p238).

No constant appeared explicitly here, as Pratt had carefully defined his units so that the coefficient of proportionality was 1. Maybe his aim was to convey clearly the explicit mathematical form without confusing the issue with physical realism.

Real life experimenters were denied this luxury, as shown clearly in Frances Bailey and George Biddell Airy's measurement of the density of the earth in 1843 using a torsion balance:

> The unit which we shall adopt is, the mass which at London weighs one grain: and for the attraction which it produces at distance 1 .... we shall multiply it by the modulus $k$.

$$k = \frac{2\pi^2.F.h}{E.i} \times \frac{s-s\prime}{T^2}$$

(where $F$, $h$, $E$, and $i$ are known instrumental parameters, $s$, $s'$ are the observed equilibrium points of the balance with the deflecting masses in two different positions, and $T$ is the observed period of oscillation) (Airy 1843 pp103, 109).

I find Airy's use of a 'modulus $k$' illuminating. A 'modulus' expresses proportionality but implies no commitment to constancy. 'Young's modulus', *E*, for example, shows that the relationship between the elastic deformation of a wire and a small, applied straining force is linear, but *E* varies from one wire to the next. Airy's choice of '$k$' for the modulus can be traced back at least to Legendre's memoir on elliptic transcendentals (Legendre 1792) – evidence for its origin in mathematical questions that ultimately derived from astronomical calculations.

Airy's caution is understandable, He was heavily involved in debates over geodesy, and what local variations in the Earth's surface gravitational acceleration (customarily signified by *g*) could reveal about gravitation. All methods of measuring MDE and putting a scale on the universe relied on an assumption that gravitation was, indeed, universal and that the same law held between parts of the Earth as between planets. Indeed, measurement of the shape and surface gravity of the Earth was the *only* way of testing that Newton's law was applicable at microscopic distances. But geodesic measurements were still far from conclusive on whether Newton's law did actually hold at terrestrial distances (Smith 2014, Ohnesorge 2024). Only once measurements began to converge, and detailed geodesic models of the shape of the Earth began to stabilise towards the end of the century, could confidence in the validity of a constant (rather than a modulus) be entertained.

By 1873, though, the constant in gravitation experiments was beginning to acquire a wider and more fundamental physical meaning. For the first time a publication gave its measurement as an objective: Alfred Cornu and Jean-Baptistin Baille's 'Détermination nouvelle de la constante de l'attraction et de la densité moyenne de la terre' (Cornu & Baille 1873). They used *f* for their constant, coupling it with the question of the mean density of the earth; their ultimate research question was the density of the planets. Eleven years later, in 1884, came the first paper that emphatically foregrounded the gravitational constant, now finally denoted by $G$, as its aim: Arthur König and Franz Richarz's *Eine neue Methode zur Bestimmung der Gravitationsconstante* (König & Richarz 1884).

My sketch of the historical buildup to the debate between Poynting and Boys might suggest that Poynting was just being a conservative reactionary in not accepting the gravitational constant interpretation of his experiments. After all, having just written a history of gravitation experiments, he was conscious of working within this historical sequence of measurements. He also agreed with Boys that these were tending to a constant value for MDE and $G$. In a slightly wider context, the challenge that geodetic experiments had presented to the universality and mutuality of Newton's law of gravitation, was near resolution through an iterative series of analyses of systematic discrepancies between measurements and refinement to the model of the shape of the earth (Ohnesorge, 2024). And Boys' experiment had extended the range of laboratory measurements of gravitational attraction down to a few centimetres.

So what was Poynting's problem in referring to MDE rather than $G$? Comparing their writings we find that the mysterious grandeur of $G$ was what appealed for Boys:

> '*G*... represents that mighty principle under the influence of which every star, planet, and satellite in the universe pursues its allotted course.... Unlike any other known physical influence, it is independent of medium, it knows no refraction, it cannot cast a shadow. It is a mysterious power, which no man can explain .... When will a St. Paul arise able to declare it unto us? Or is gravitation, like life, a mystery that can never be solved?' (Boys 1894 p355)

This, though, was the very essence of Poynting's objection. Gravity *was* 'unlike any other known physical influence,' and declaring $G$ as a universal constant precluded further probing into its nature; it was a scientific dead-end.

Poynting held strongly that physical laws were purely descriptive. In 1899 he imagined a historical evolution through which physicists' concepts of 'physical law' had passed to reach what he believed was a current consensus:

> 'No long time ago they were quite commonly described as the Fixed Laws of Nature, and were supposed sufficient in themselves to govern the universe. Now we can only assign to them the humble rank of mere descriptions.... The old conception of laws as self-sufficing governors of Nature was, no doubt, a survival of a much older conception of the scope of physical science. I imagine that originally man looked on himself.... Knowing that his purpose and will were followed by motions and changes in the matter about him, he thought of similar purpose and will behind all the motions and changes which he observed, however they occurred.... Taking this anthropomorphic...view, the laws he formulated were... expressions of fixed purpose.... They were commands given to matter which it must obey.
>
> These early stages in the history of thought can be traced by survivals in our language, survivals due to the ascription of moral qualities to matter. Thus gases are still sometimes said to obey or to disobey Boyle's Law as if it were an enactment for their guidance, and as if it set forth an ideal, the perfect gas, for their imitation.' (Poynting 1899 pp616-617)

Physicists in the 1890s, he believed, now knew better and arrived at descriptive laws by,

> 'finding out all likenesses... classing together all similar events, and so giving an account as concise as possible of the motions and changes observed.... Were there no similarities in events, our account of them could not rise above a mere directory.... But the similarities observed enable us ... to make... a readable book of science, with laws as the headings of the chapters.' (Poynting 1899 pp615-616)

Poynting went on to give three examples of laws which were, 'in all cases brief descriptions of observed similarities':

1. Gravitation which states that, 'to each portion of matter we can assign to it a constant – its mass – such that there is an acceleration towards it of other matter proportional to that mass divided by the square of its distance away. Or all bodies resemble each other in having this acceleration towards each other',

2. Hooke's law for a stretched wire which 'states that the behaviour of the wire is similar for all equal small pulls', [generally expressed today as $F_S = kx$ where $F_S$ is the stretching force, $x$ is the displacement of the free end of the wire as it stretches, and $k$ is a characteristic of the particular wire],
3. Joule's law for the heat appearing when a current flows in a wire in which 'all the different cases resemble each other in having $H \div C^2Rt$ constant' [where $H$ is the heat developed, $C$ is the current, $t$ is time, and $R$ is the resistance, specific to the wire] (Poynting 1899 p616).

Note that he treats all three laws with equal status. What was important to his argument was that in each law the mathematical form applied to all cases. He presented each law in a form that emphasised the constant that was specific to the particular body or wire involved, mass, $k$ and and $R$ respectively. He made no mention that for gravitation the proportionality is the same over all cases – unlike the other two examples. This provides further evidence of his reservations about $G$.

These three examples exemplify Poynting's view that alongside the change in concept of physical law had gone a change in what counted as a physical explanation. No longer were underlying causes sought:

> 'On the purely physical or descriptive view, the idea of cause is quite out of place. In description we are solely concerned with the 'how' of things, and their 'why' we purposely leave out of account.' (Poynting 1899 p617)

This was the crucial point for his views on the gravitational constant, for,

> 'We explain an event not when we know 'why' it happened, but when we show 'how' it is like something else happening elsewhere or otherwhen [sic] – when, in fact, we can include it as a case described by some law already set forth.' (Poynting 1899 p617)

If physicists conceded that gravity was unlike anything else, then they cut off the possibility of such physical explanations, and of any understanding through analogy or assimilation into wider laws. The most obvious analogy to the law of gravitation was the inverse square law of electrostatics. Electrical science had advanced significantly in the previous 50 years, partly through the close attention and careful experiments that had shown that the medium and state of charged bodies *did* exert an influence on the electrostatic force – the inverse square relation held, but the constant had to be modified to recognise the medium, and limitations put on the applicability of the law. Nothing like this was yet known for gravity but Poynting held that the physicist's task should be seeking ways in which $G$ could be forced to vary.

Thus Poynting, and others who thought as he did, embarked on laboratory trials. For example, Poynting attempted to find a difference in the attraction between quartz crystals with the crystalline axes in different orientations (Poynting & Gray 1898), while Mackenzie (1895) compared the attractions of crystalline and isotropic masses. Poynting & Phillips (1904) and Southerns (1907) investigated temperature effects on gravity, and Braun (1897) tested the influence of low pressures. Austin & Thwing (1887), Cremieu (1907), Erismann (1908) all compared gravitational forces in different media, including radioactive ones (Geigel 1903). Bauer (1907) looked for variations in the apparent mass of a magnet in various magnetic fields, and Sanford and Ray (1897) checked for possible changes of weight in chemical reactions.

However, by 1900 when Poynting addressed the Royal Institution on ‘Recent Studies in Gravitation’, none of these attempts had borne fruit. Nor were they to in the succeeding years. Poynting concluded despondently:

> ‘... while the experiments to determine *G* are converging on the same value, the attempts to show that, under certain conditions, it may not be constant, have resulted so far in failure all along the line... Perhaps we have been led astray by false analogies in some of our questions. Some of the qualities we have sought and failed to find, qualities which characterise electric and magnetic forces, may be due to the polarity, the + and -, which we ascribe to poles and charges, and which have no counterpart in mass. But this unlikeness, this independence of gravitation of any quality but mass, bars the way to any explanation of its nature... Gravitation still stands alone.’ (Poynting 1900 p294).

## 4 Discussion

In his conclusion, quoted above, Poynting suggests explicitly that not recognising and questioning the assumptions of the tradition one is working in – in this case a tradition of reasoning by analogy from electrostatics to gravity – may be a barrier to development and lead to false conclusions. We have seen in the case of Maxwell and Thomson that such lack of awareness did, indeed, lead to confusion and conclusions that were only weakly founded mathematically. But whereas Maxwell and Thomson were saved from false physical conclusions by (generally unacknowledged) experimental induction, Poynting respected the sequence of physical measurements that stood in the way of a fuller explanation of gravity. He pinpointed the analogy of mathematical form between electrostatics and gravitation as the origin of the problem, but was unwilling or unable to step outside this tradition. It was only when Einstein did step outside, and sought a new mathematics, that understanding of gravity moved on.

The difficulty of freeing oneself from tradition is further shown by Maxwell & McAlister, and Cavendish, apparently not realising that they were using instruments that relied on the very law they were testing. Similar issues arose in geodetic measurements to measure the ellipticity of the earth, where survey theodolites were oriented on the assumption of an ellipsoidal shape, although in this case measurers were aware of the problem. In both cases early measurements ultimately formed the beginning of iterative sequences that more consciously relied on the model they were testing as a first approximation in refining the model and the measurement. Cavendish and Maxwell’s deductive null method has formed the basis of increasingly precise demonstrations of the electrostatic law ever since, notably by Plimpton and Lawton (1936), Bartlett et. al. (1970), Williams et. al. (1971) and Crandall (1983). In his critique of Cavendish and Maxwell, Laymon (1994) included these twentieth century experiments, positing a developmental process in which successive experimenters have assumed, in their sensitivity analysis, the inverse square law to at least the precision of their predecessors. Chang (2007) termed a similar process of iterative improvement of model-based measurements, ‘epistemic iteration’. In analysing the way in which geodesists ultimately demonstrated the validity of mutual interaction in Newton’s law and arrived at a detailed knowledge of the shape of the earth, Ohnesorge (2024, p152) has added the use of ‘residual knowledge’ to epistemic iteration – novel ideas drawn from the investigation of residual errors between current measurement and model. Such processes are historical in the sense that they take place through time as measurements and models succeed one another. They are frequently observed in fields where theoretical entities (such as $G$) are measured indirectly, based on mathematical models.

The course of gravitation experiments through the 20th and early 21st centuries has shown that Poynting and Boys were over-sanguine in their belief that measurements of MDE and $G$ were converging to a consistent and stable value. Its measurement precision has increased by only about two orders of magnitude since Cavendish's 1798 experiment, and values of $G$ have not converged. For this reason, $G$ does not form one of the bases of the new definitions of S.I. units (Quinn & Speake 2014). The current accepted value 6.674 30(15) x $10^{-11}$ $m^3kg^{-1}s^{-2}$ has a relative standard uncertainty of 2.2 x $10^{-5}$ reflecting a situation where the spread of recent results (since 1982) is ten times the uncertainty in any single result (Morh et. al. 2025). A major international conference in 2014 and top level working groups since, have failed to resolve the discrepancies. Current efforts focus on proliferating the number of independent methods used to measure G at the same location in an attempt to identify systematic errors (Wu et al 2019). Alternatively, unknown physics may be affecting the results, but the very uncertainty in $G$ is a barrier to testing hypothesised effects (Wu et al 2019). The model of historical progress suggested by Chang and Ohnesorge of epistemic iteration plus residual knowledge, requires coherent results as a pre-condition, and currently $G$ measurements do not satisfy this condition.

The gravitation case study provides an example of the way in which mathematical form may have impact on physicists' concepts and research questions about the natural world. The nature of measurement changed alongside the mathematical form of the law of gravitation. Eighteenth-century workers proceeded directly to MDE. The question of a gravitational constant did not arise, and they did not even need to measure the attraction between the two smaller masses directly in terms of their mass – one of the masses falls out of the ratio equations (Figure 4a). This left Newton's assertion of the mutuality of the attraction ($m_1m_2$) untested, and it was taken up by geodesists investigating the shape of the earth instead. The introduction of what I have called the explicit form of the law, introduced a new mathematical entity into physics, initially as a 'modulus' that satisfied dimensionality requirements. Through the 19th century what started as a mathematical entity became an object of study in its own right, a universal constant to which physical significance and meaning was attached. The purpose and the calculational procedure of experiments changed, prioritising determination of $G$. Even Poynting, who felt that the *universal* constant concept was not useful, obtained $G$ first as a step in the calculation of MDE.

The historical processes of interaction between mathematics and physical experiment discussed above are seen throughout mathematical modelling of the natural world. But both case studies discussed here show a different type of history also: these men explicitly invoked historical cases in support of their views. But there the similarity ends. Poynting made clear that his history was supposition – a reconstruction that seemed plausible to him. Although, as he pointed out, traces of earlier concepts of physical law are apparent in our language (even today we refer to 'ideal' gases), he made no attempt to put a chronology on the change or attribute it to specific people. As far as one can infer his purpose, it was simply to emphasise his interpretation of laws as purely descriptive by pointing a contrast to a supposed former idea of laws as governors of Nature. The case of Thomson and Maxwell was very different. They attributed developments to specific people, and wrote others out of their narrative, generally showing a nationalistic preference for British workers. They misrepresented Faraday's work, apparently knowingly, leveraging his reputation to support their cause. The history implicit in their account may seem careless at the very least, possibly even unethical, although it is arguable that the cause they were pursuing – mathematical approaches to physics – was benefiting society in the shape of improved telegraph communications and technology. They provide a cautionary example of the way in which audiences may be manipulated and interest groups bolster their authority, especially if such narratives are provided by respected figures.

Both these case studies have revealed entanglements between mathematics, physics, and historical process that are often inevitable, but that are not evident in the quick conventional answers to why Maxwell repeated Cavendish's experiment or Poynting preferred to measure the mean density of the earth. In doing so they have surfaced some of the risks of lack of awareness of the tradition within which one is working or of uncritical acceptance of traditions presented.